\documentclass[fleqn,twoside]{article}
\usepackage{amssymb}
\usepackage[headings]{espcrc2}

\readRCS
$Id: espcrc2.tex,v 1.2 2004/02/24 11:22:11 spepping Exp $
\usepackage{graphicx}
\usepackage[figuresright]{rotating}

\newcommand{\AmS}{{\protect\the\textfont2
  A\kern-.1667em\lower.5ex\hbox{M}\kern-.125emS}}

\def\mpipi{M_{\pi\pi}}

\title{Scalar mesons: in search of the lightest glueball}

\author{Wolfgang Ochs\address[MCSD]{Max-Planck-Institut f\"ur Physik,
        F\"ohringer Ring 6, D-80805 M\"unchen, Germany}}

\begin{document}

\begin{abstract}
According to the QCD expectations the lightest glueball should be a scalar
particle ($J^{PC}=0^{++}$). Different scenarios have been
considered for a classification of these states but -- despite considerable
progress in recent years -- the experimental basis for various parameters
is still rather weak. We present a new analysis
of the elastic and charge exchange $\pi\pi$ scattering between 1000 and 1800
MeV. A unique solution is selected which shows clear evidence for
$f_0(1500)$
and a broad state ($\sigma$ or ``$f_0(1000)$''), but there is no evidence for
$f_0(1370)$ at a level of $\gtrsim 10$\% branching ratio to $\pi\pi$. 
Arguments in favour of the broad state to be a glueball are summarized.
\vspace{1pc}
\end{abstract}

\maketitle

\section{EXPECTATION FOR GLUEBALLS IN QCD}
The existence of glueballs has been considered from the very beginning 
as characteristic prediction of QCD because of the self-interaction of
gluons; early investigations suggested different scenarios \cite{fm}.
Today, quantitative results are available from\\
{\it (1) Lattice QCD} \\
Computations in quenched approximation locate the lightest
glueball in the $0^{++}$ channel in the mass range 1400-1800 MeV
(see review \cite{bali}).\footnote{After the conference the result from
an unquenched calculation \cite{unquenched} has been reported:
gluonic and $q\bar q$ states mix and the mass of the light flavour
singlet has dropped from 1600 to 1000 MeV.}\\ 
{\it (2) QCD sum rules}\\
Detailed results on glueballs together with a scenario for the $q\bar q$ sector
have been obtained by Narison \cite{narison}. The glueball around 1500 MeV
is reproduced, in addition a lighter gluonic state around 1000 MeV is
required with strong decay into $\pi\pi$ and a large width mixing with
the nearby $q\bar q$ state.  Other analyses for the gluonic sector
find similar results with two states \cite{steele} or only one state in the
region around 1250 MeV \cite{forkel}.

\section{INTERPRETATIONS OF THE OBSERVED SPECTRUM}
From these computations one would expect the lightest glueball 
in the scalar sector in the mass
range up to about 1800 MeV. In this search it 
is necessary to identify at the same time the
$q\bar q$ (or possibly the $qq\bar q\bar q$) nonets and a possible mixing
of states.  
The Particle Data Group lists the following scalar particles \cite{pdg}:\\
I=0: $f_0$(600) ($\sigma$), $f_0(980)$, $f_0(1370)$, $f_0(1500)$,
\hspace*{8mm} $f_0(1710)$\ldots\\
I=$\frac{1}{2}$: $K^*_0$(800) ($\kappa$) (?), $K^*_0(1430)$, $K^*(1950) (?)$\ldots\\
I=1:  $a_0(980)$, $a_0(1450)$\ldots\\

There are different scenarios for the interpretation of this spectrum
and we describe shortly two variants, for more details and references, see,
for example, Ref. \cite{mobdec}.\\
{\it Scenario A}:\\
A light nonet is formed with the states $\sigma(600),\kappa(800), a_0(980),
f_0(980)$, either with $q\bar q$ or $qq\bar q\bar q$ composition.
Then, a heavier nonet can  be built with $a_0(1450),\ K^*_0(1430)$ and 
two isoscalars which mix with the bare glueball into the observed 
$f_0(1370), f_0(1500), f_0(1710)$ (for an early Ref., see \cite{ac}).\\
 {\it Scenario B}:\\
The lightest members of the nonet are $f_0(980)$ and $a_0(980)$ (or
$a_0(1450)$), furthermore $K^*_0(1430)$ and $f_0(1500)$ (or $f_0(1370)$).
The glueball is considered a very broad object with mass in the region 1000
MeV \cite{mo} - 1500 MeV \cite{anis} and width of the order of mass.

\begin{figure*}[t]
\begin{tabular}{@{}lll}
\includegraphics*[angle=-90,width=5cm,bbllx=3cm,bblly=1.5cm,bburx=19.5cm,
bbury=19.5cm]{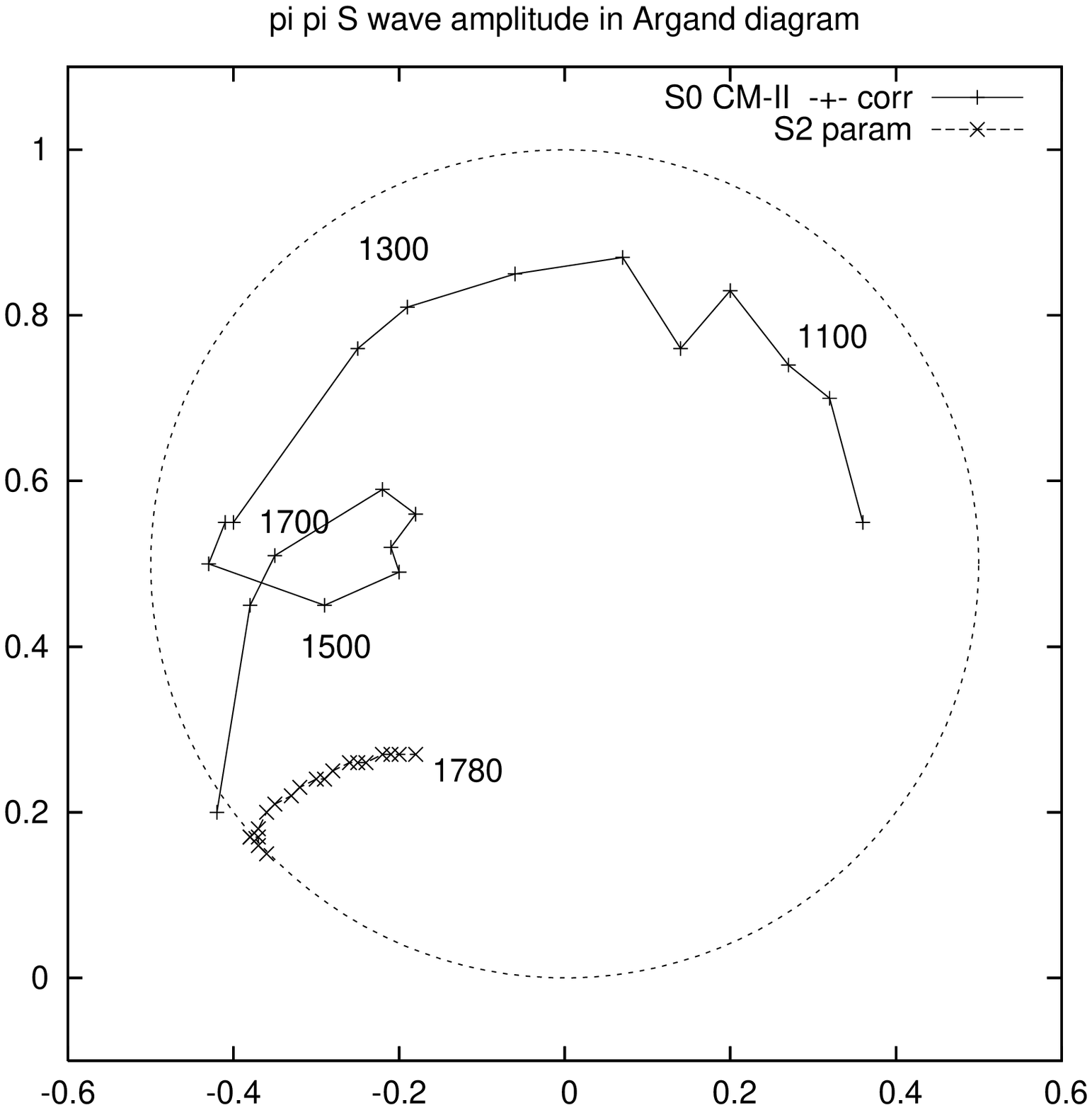},
\includegraphics*[angle=-90,width=5cm,bbllx=3cm,bblly=1.5cm,bburx=19.5cm,%
bbury=19.5cm]{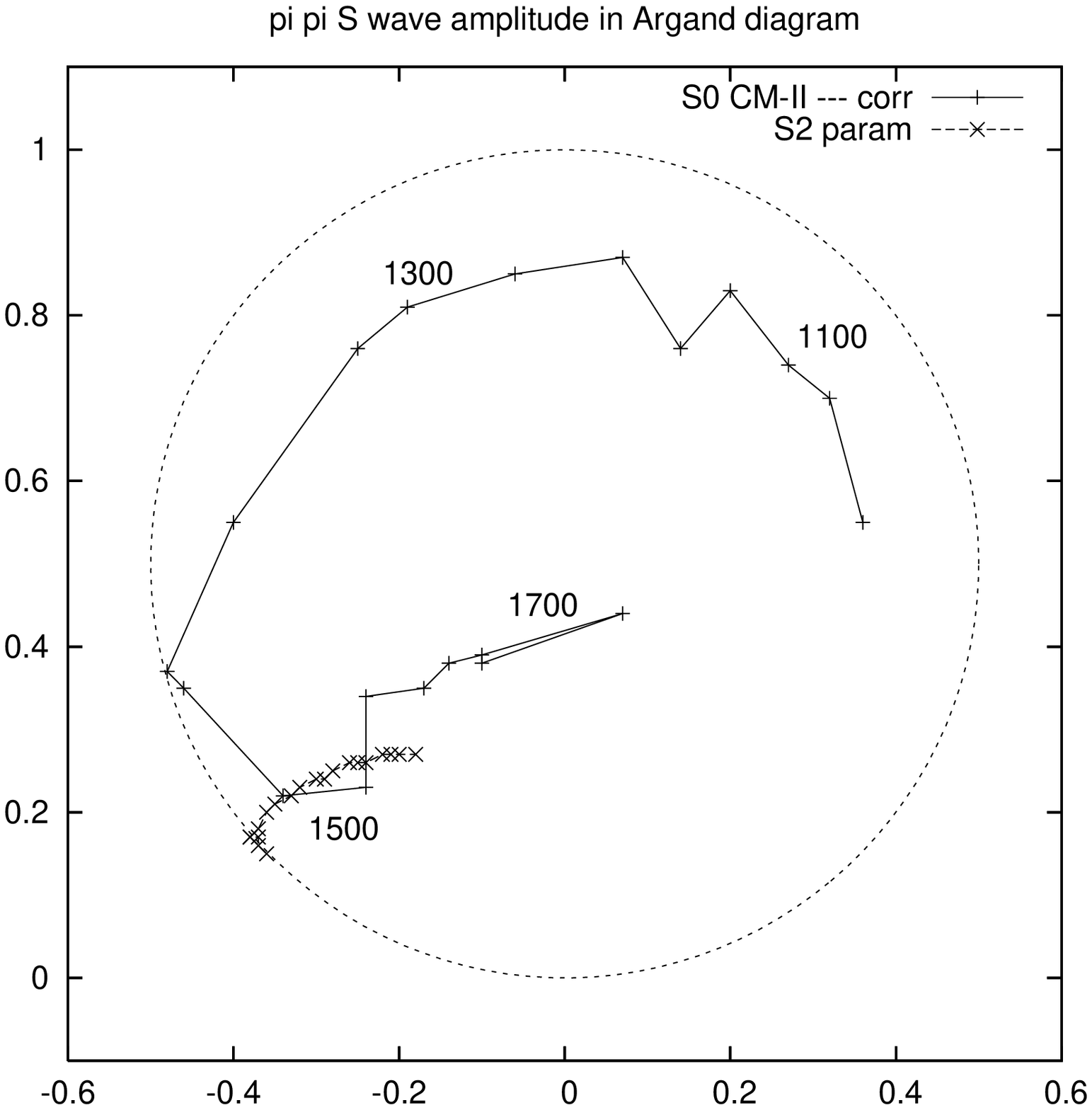},
\includegraphics*[angle=-90,width=5cm,bbllx=3cm,bblly=1.5cm,bburx=19.5cm,%
bbury=19.5cm]{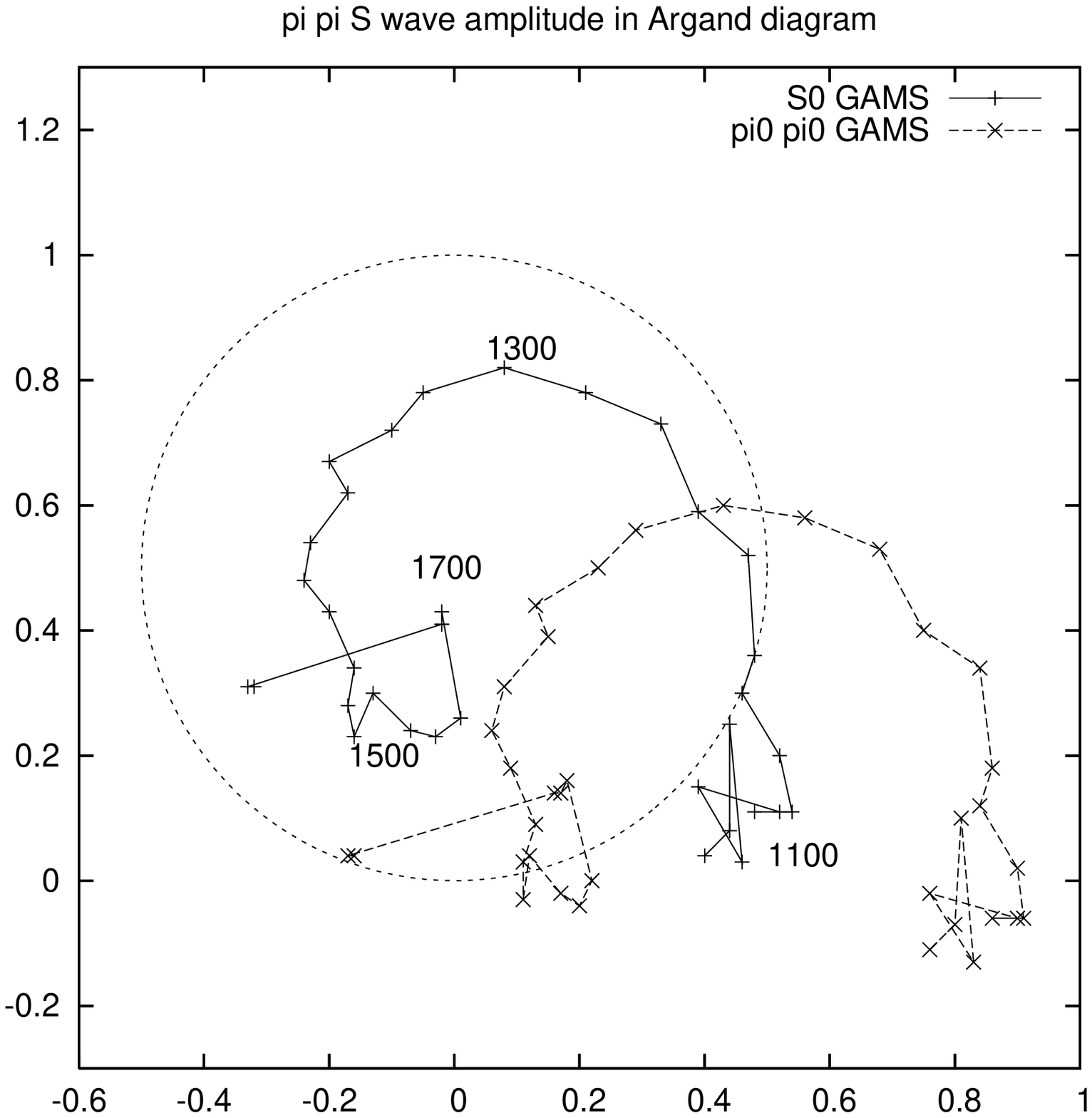} \\
\end{tabular}
\caption{Argand diagrams for $\pi\pi$ $S_I$ waves: $S_0$ solutions  $-+-$ and
$---$ from elastic $\pi^+\pi^-$ (CM-II),
corrected for new I=2 results, and $S_{00}$ amplitude for 
 $\pi^+\pi^-\to \pi^0\pi^0$ (GAMS) and component $S_0$ thereof.
Also shown on the left is the amplitude $S_2$.}
\label{fig:swave00}
\end{figure*}

The scheme for light scalars proposed by us \cite{mo} follows Scenario B.
The isoscalar partners $f_0(980)$ and $f_0(1500)$ are ``non-ideally'' mixed
near an $SU(3)$ flavour singlet and octet respectively, just like $\eta'$ and
$\eta$ but with opposite mass ordering as also suggested in \cite{klempt}.
This proposal is consistent with various observed decay rates; 
especially, the relative phases
between the $s\bar s$ components in $f_0(980),\ f_0(1500)$ follow from
interference effects \cite{mo,momontp}.
In addition, this multiplet fulfils the
Gell-Mann Okubo mass formula. 
Ultimately, we hope that the correct choice of the 
$q\bar q$ multiplet will be established by the verification of 
flavour symmetry relations, 
such as those suggested for charmless $B$-decays \cite{mobdec}. 

We don't
discuss these arguments here any further but try to elucidate the properties of
$f_0(1370)$ -- which has an important impact on the 
classification schemes -- and
the broad object, our glueball candidate. Phenomenologically, one observes
in the elastic $S$ wave $\pi\pi$ scattering cross section 
a broad bump modified by dips at the masses
of $f_0(980)$ and $ f_0(1500)$ (the ``red dragon''). All resonance 
fits to this
phenomenon, first observed by the CERN Munich collaboration (CM-I
\cite{cm}), revealed a pole in the $T$ matrix corresponding to a state 
with $M\gtrsim 1000$ MeV and comparable width 
(also called $f_0(1000)$ \cite{mp}). 
This object may be identical
with the broad $\sigma$ particle.

\section{NEW ANALYSIS OF $\pi\pi$ SCATTERING FOR $M_{\pi\pi}\gtrsim 1000$ MEV.}

The first analysis of CM-I data \cite{cm} on $\pi^+\pi^-\to \pi^+\pi^-$
in the range $600 \leq
M_{\pi\pi}\leq 1800$ MeV was based on an energy dependent $K$- matrix fit
followed by a bin-by-bin energy-independent analysis. Ambiguities expected
on general grounds have not been excluded above 1000 MeV. Subsequently,  they  
have been
classified into 4 types by Estabrooks and Martin \cite{em} and in CM-II
\cite{cm2} according to the ``Barrelet zero'' systematics. 
Additional information has been added by the CERN-Krakow-Munich (CKM) 
experiment with polarised target \cite{ckm,klr}. 

\begin{figure*}[t]
\begin{tabular}{@{}lll}
\includegraphics*[angle=-90,width=5cm,bbllx=3cm,bblly=1.5cm,bburx=19.5cm,
bbury=19.5cm]{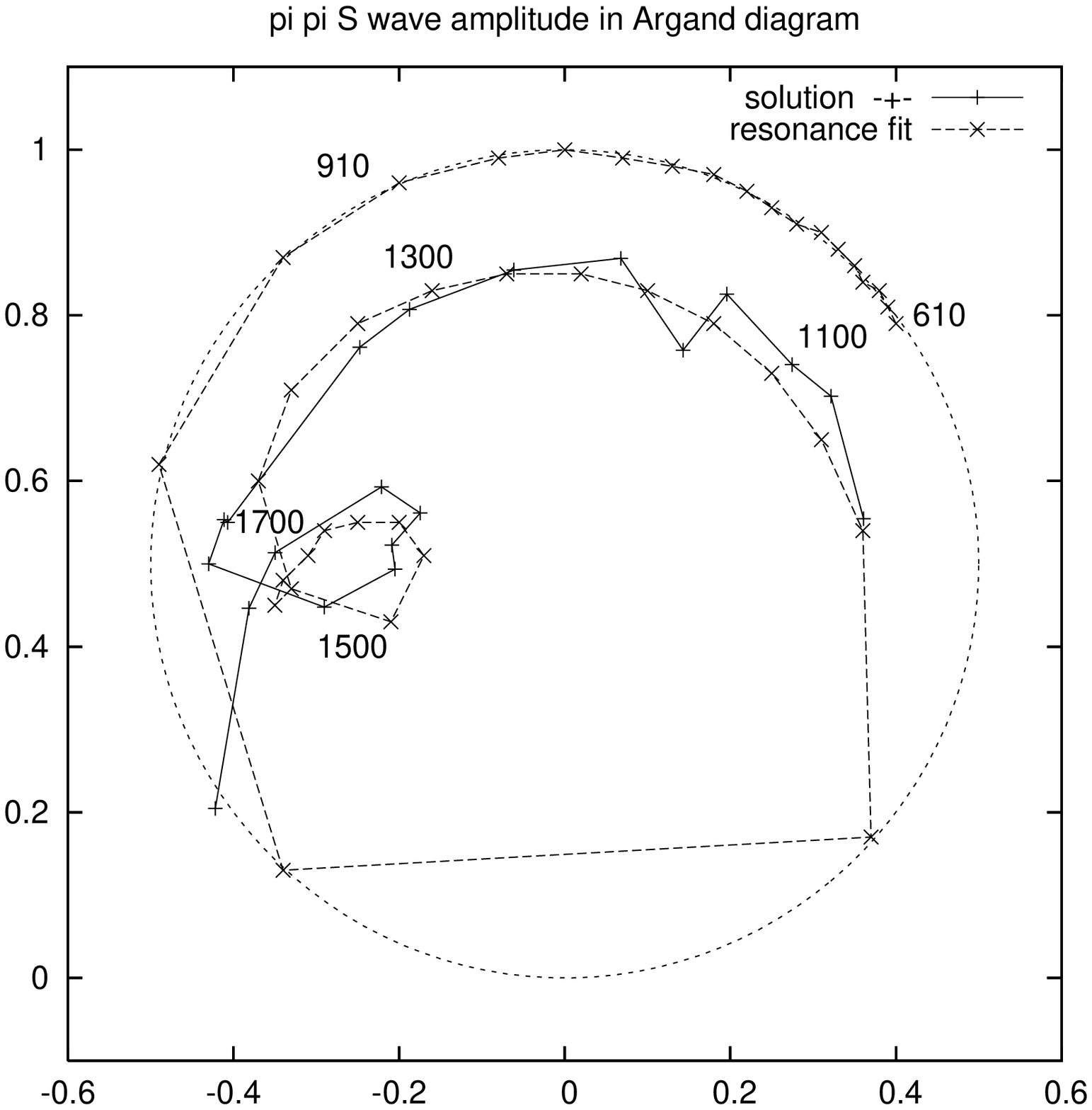},
\includegraphics*[angle=-90,width=5cm,bbllx=3cm,bblly=1.5cm,bburx=19.5cm,%
bbury=19.5cm]{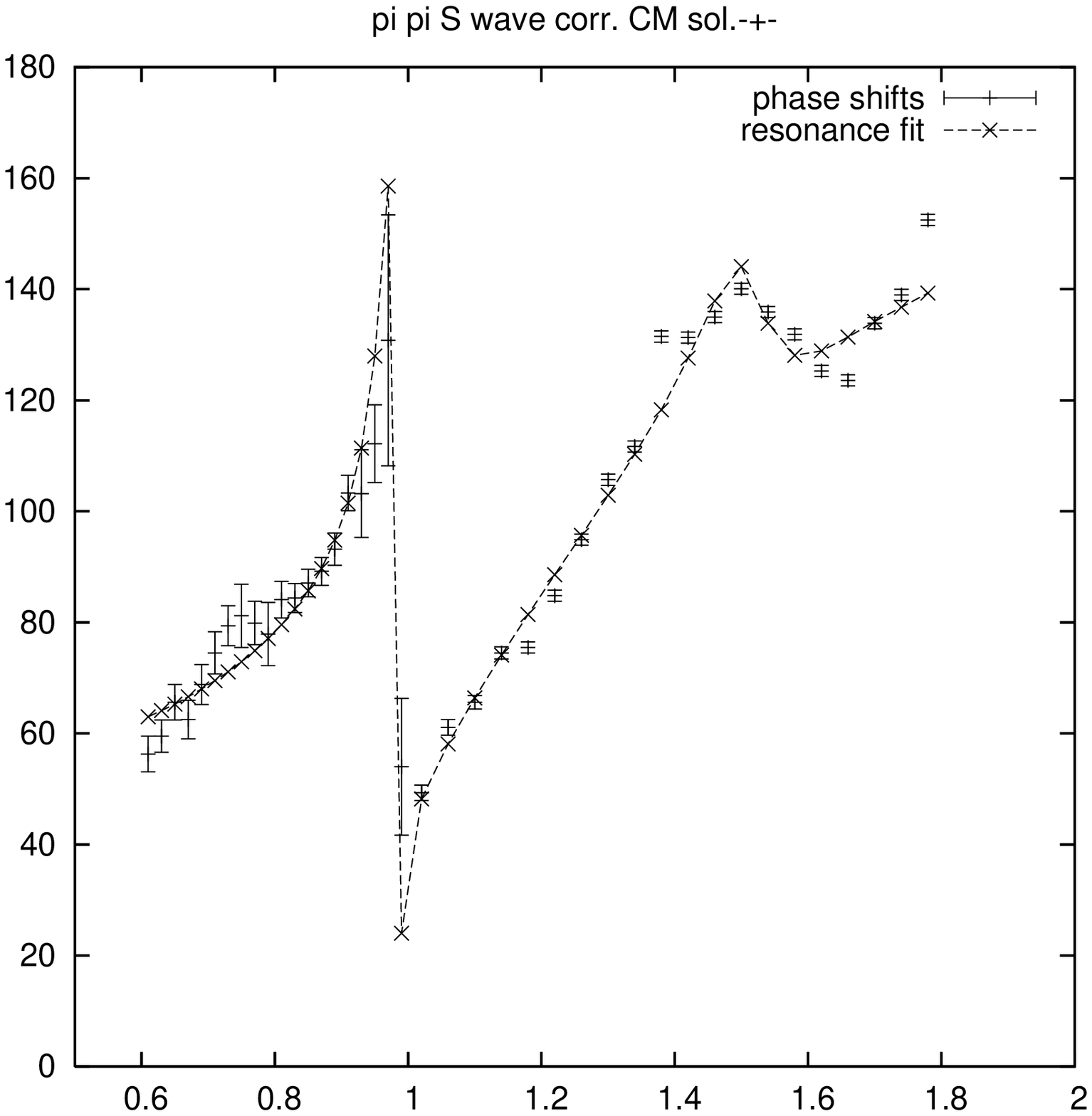},
\includegraphics*[angle=-90,width=5cm,bbllx=3cm,bblly=1.5cm,bburx=19.5cm,%
bbury=19.5cm]{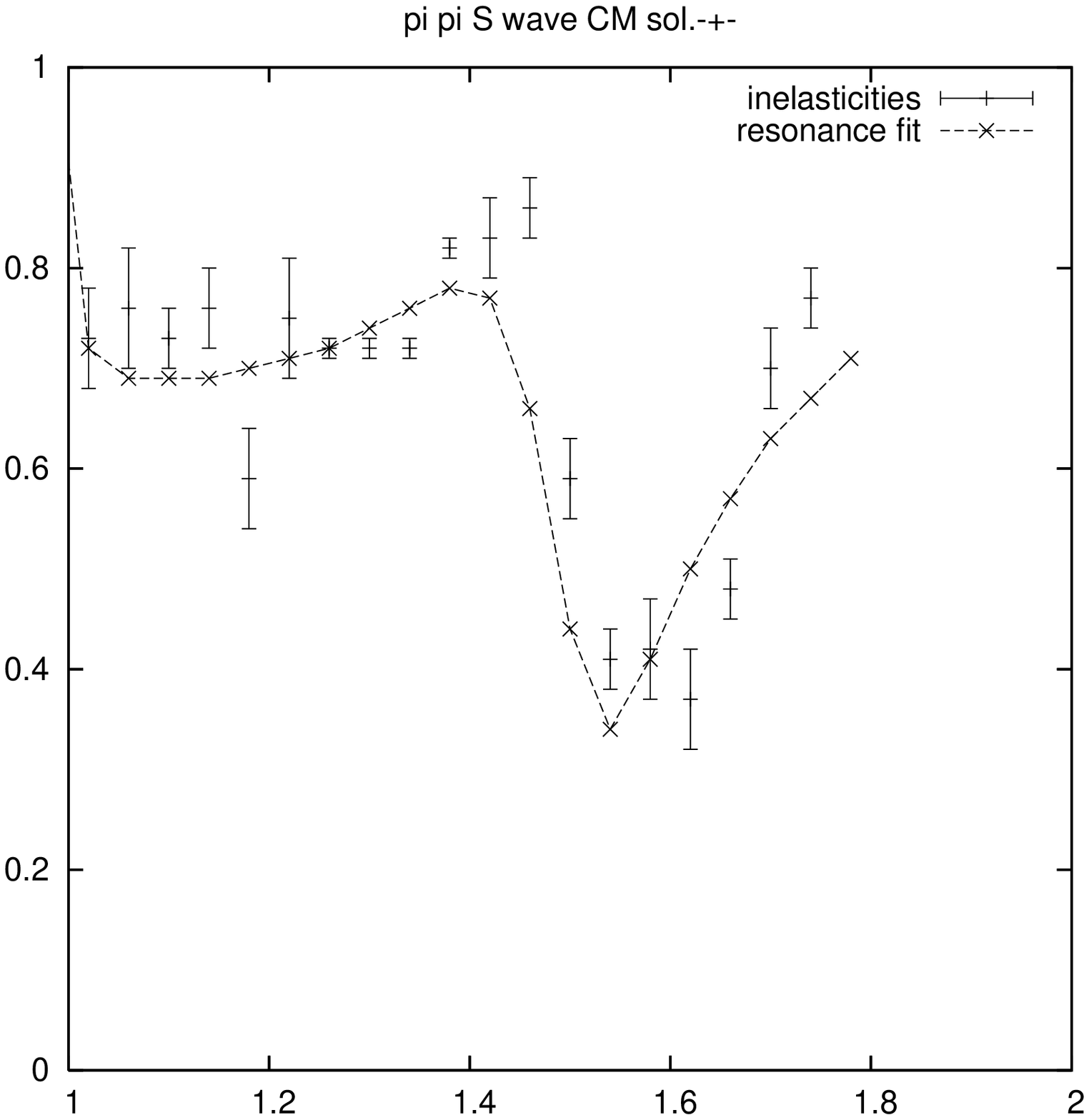} \\
\end{tabular}
\caption{Resonance fit Eq. (1) in comparison with data (CM-I/II): 
Argand diagram for corrected $S_0$ wave,
phase $\delta_0^0$ and inelasticity $\eta^0_0$.}
\label{fig:resonances}
\end{figure*}

\subsection{Selection of the physical solution}
A unique solution for $\mpipi<1400$ MeV has been
suggested recently by Kami\'nski, Pel\'aez and Yndur\'ain \cite{kpy} 
who found good agreement for the $S$ wave 
phases from the three above mentioned determinations up to 1400
MeV: CM-I, CM-II (Sols. $---$ or, equivalently here $-+-$, or Sols. A,C 
\cite{em}), and
CKM, but not Sols. $+--$ and $++-$ (B,D). 
 
We can obtain a further selection among the remaining ambiguities above 1400 MeV
by comparing with 
the reaction $\pi^+\pi^-\to \pi^0\pi^0$ with its different ambiguity
structure. Such data have been
obtained from $\pi^- p$ collisions at 100 GeV/c (GAMS 
\cite{gams}) and at 18.3 GeV  (E852 \cite{bnl}). For our
study we use the GAMS data which extend to  masses $\mpipi>1400$ MeV with good
accuracy. 

We also take into account the $I=2$ amplitude $S_2$. 
For the mass range $1000\lesssim \mpipi \lesssim 1800$ MeV we represent the data
\cite{durusoy,cohen} on phases and inelasticities by the
parametrizations 
$\delta_2^0=-25^\circ+25^\circ(\mpipi-1.3)^2$ and 
$\eta^0_2={\rm Min}[1.0-(\mpipi-1.2)+0.5(\mpipi-1.2)^2,1.0] $ (masses in GeV). 
The errors on the inelasticity at the higher energies are rather large 
$ \eta^0_2\sim 0.5\pm0.2$.
 
For further analysis of elastic $\pi^+\pi^-$ scattering we take the CM-II results 
which use the full correlation
matrix in the determination of the amplitudes and therefore have small
statistical errors, but the results in Ref. \cite{em} are qualitatively 
similar. 
As CM-II used an older purely elastic $S_2$ wave
we recalculated $S_0$ from the measured $\pi^+\pi^-$ amplitude 
assuming $S_{+-}=(S_0+(1/2)S_2)_{old}=(S_0+(1/2)S_2)_{new}$. The remaining 
two $I=0$ amplitudes  $S_0$ ($-+-$ and $---$)
corrected in this way are shown in
in Fig. \ref{fig:swave00} together with the $I=2$ wave. 

Rightmost in Fig. \ref{fig:swave00} is shown
the amplitude $S_{00}$ for the process $\pi^+\pi^-\to \pi^0\pi^0$
which we extracted from the magnitude $|S_{00}|^2$, the phase
$|\Phi_S-\Phi_{D_0}|$ and the Breit Wigner phase of $f_2(1270)$ in the 
$D$ wave as determined from GAMS data \cite{gams}. The $I=0$ amplitude $S_0$ is obtained from $S_{00}=S_0-S_2$ 
and is shown in the same diagram.
A second solution from the unresolved sign of $\Phi_S-\Phi_{D_0}$
 is rejected by the unitarity requirement.

The amplitude $S_0$ so obtained shows a circular motion from 1000 to 1740
MeV with a smaller circle superimposed above 1400 MeV. This behaviour is
only consistent at the qualitative level with the  $\pi^+\pi^-$ solution $-+-$ 
which we therefore select as the physical solution. There are some remaining
discrepencies between the results on $S_0$ from both processes. These we
attach to systematic errors in the determination of the overall phase from
leading resonances, from truncation of higher partial waves and the
uncertainty in the $I=2$ amplitude. These uncertainties over large mass
scales should not affect
the nature of local resonance phenomena like $f_0(1500)$ or  $f_0(1370)$.

It is nevertheless satisfactory that agreement within these limitations is
obtained. It shows the production of $f_0(1500)$, the small circle, 
with parameters roughly according to PDG (M=1500 MeV, $\Gamma$=109 MeV)
and with branching ratio into $\pi\pi$ of $x_{\pi\pi}\sim0.349$, 
whereas we estimate from the depth of the circle  $x_{\pi\pi}\sim0.25$. 
The polarised target 
CKM data on $|S|^2$ \cite{ckm}
which are determined in a model independent way agree better with CM-II for
$M\lesssim 1400$ MeV and better with GAMS for $M\gtrsim 1400$ MeV
with its higher inelasticity.

\begin{figure}[t]
\begin{center}
\includegraphics*[angle=-90,width=5cm,bbllx=3cm,bblly=1.5cm,bburx=19.5cm,
bbury=19.5cm]{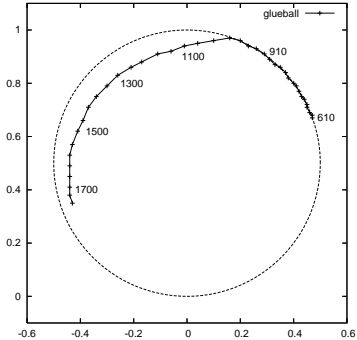}
\end{center}
\caption{Argand diagram for the broad component ($f_0(1100)/\sigma$),
the hypothetical glueball,
in our resonance fit of Fig. 2, to CM-II data. GAMS data would suggest
larger inelasticity at higher masses.}
\label{fig:glueball}
\end{figure}

\subsection{Resonance fit to the $I=0$ $S$ wave}

Next we look for resonances in the solution $S_0(-+-)$ of CM-II
(corrected). We represent the $S$ matrix for the 3 channels ($\pi\pi,K\bar
K,4\pi$) as product of 3 matrices for resonances
$S_R=1+2iT_R$

\begin{eqnarray}
S&=&S_{f_0(980)}S_{f_0(1500)}S_{\rm broad}\\
T_R&=& [M_0^2-\mpipi^2-i(\rho_1 g_1^2 +\rho_2 g_2^2+ \rho_3 g_3^2)]^{-1}
   \nonumber\\
   & &   \phantom{xxx}\times\rho^{\frac{1}{2}T} (g_ig_j)\rho^\frac{1}{2}
    \label{smatrix}
\end{eqnarray}
where $\rho_i=2k_i/\sqrt{s}$.
We fit these formulae to the data from CM-II for $\mpipi>1000$ MeV and the
phases from CM-I in $600<\mpipi<1000$ MeV for definiteness. 
The details will be given elsewhere. As can be seen
in Fig. \ref{fig:resonances} the three resonances give a good overall
description of the data. 

The $\pi\pi$ component of $S_{\rm broad}$ in (1) is shown separately in Fig.
\ref{fig:glueball}. It corresponds to a mass parameter in (2) of $M_0$=1100
MeV and a total width of similar size. The parametrization (2) with constant $g_i$ 
has not the
threshold behaviour expected from chiral theory. The Adler zero can be
enforced by multiplication of the couplings with 
$(s-m_\pi^2/2)/(s-s_A)$. We found that this modification has little effect
in the mass
range considered here ($\mpipi>600$) MeV, but one may
expect that the pole position in the amplitude has shifted 
from 1100 MeV towards a lower value. Therefore the $f_0(1000)$
and the $\sigma(600)$ effects are presumably the same phenomena. 
The dependence of the pole position on the actual line shape can be
seen in a simple unitary model amplitude with Adler zero, 
communicated to us  by H. Zheng \cite{zheng} ($m_\pi\to 0$):
$T=\rho s g^2/(M^2-s-i\rho s g^2)$ with pole at $s_\sigma\sim M^2/(1+ig^2)$.
In the extreme cases, for
small coupling the pole is near $s_\sigma=M^2\ (\sim 1000$ MeV), whereas for
very strong coupling $g^2\gg 1$  Re $s_\sigma\to 0$, Im $s_\sigma \gg $ 
Re $s_\sigma$.

We note that there is no evidence for 
$f_0(1370)$ from our fitting. This state 
would show up as a second circle in the Argand diagram
and as another dip in $\eta^0_0$ at the
corresponding mass. If we allow for such a state in the fit
the mass tends towards 1500 MeV and the width to decrease such that the
additional circle fits
into the first one. We therefore exclude an additional 
state with branching ratio  $x_{\pi\pi}\equiv B(f_0(1370)\to \pi\pi)\gtrsim
0.1$ near 1370 MeV.


Results are also available on the inelastic processes 
$\pi\pi\to \eta\eta,K\bar K$
where the reconstructed Argand diagrams \cite{mo} show one extra circle
above background. Explicit fits \cite{estabrooks,cohenkk} have denied a second
narrow resonant state beyond $f_0(1500)$, contrary to the recent study
\cite{buggpipi}, so this situation has to be clarified. The 
main evidence for $f_0(1370)$ comes
from fits to 3-body final states in low energy $p\bar p$ annihilation 
and $J/\psi$
decays, also central $pp$ production of the decays $f_0\to \pi\pi,K\bar K$. 
None of these studies have presented a full 2-dim 
energy-independent phase shift analysis showing the existence of a second
circle. Such studies 
are essential in processes with large
background from a broad resonance and considerable
$I=2$ components. 

\section{Glueball interpretation of ``$f_0(1000)$''}
We see the following arguments in favour of the broad state to be a glueball.\\
1. In our interpretation of the spectrum the broad state, presumably
identical to $\sigma(600)$, does not belong to
a nonet. In particular, we would not add this effect centered around 1 GeV,
at least not entirely, to a light nonet of any kind, so it is ``left over''
in the $q\bar q$ systematics.\\
2. The state is produced in most processes considered as ``gluon rich''.
In radiative $J/\psi$ decays a broad or constant contribution has been seen 
in $K\bar K \gamma$ \cite{besgkk}, in $\sigma\sigma\gamma$ \cite{bugggss},
but not yet in $\gamma \pi\pi$ which may be difficult; furthermore in
central production, like $pp\to p\pi\pi p$ and in annihilation, like $p\bar p
\to 3\pi$ and also in $B\to K +X$ \cite{mobdec} with the
gluonic contribution $b\to s g$.\\
3. The broad state appears together with $f_0(1500)$ in channels 
$\pi\pi\to\pi\pi,K\bar
K,\eta\eta$ and $B\to K (\pi\pi),K(K\bar K)$. The signs of interference in
these channels are consistent with $f_0(1500)$ being a flavour octet (as in models
\cite{mo,klempt}) and $f_0(1000)$ being near a flavour singlet \cite{mobdec}.\\
4. Another expectation is the suppression of glueball production in
$\gamma\gamma$ processes. Based on the earlier result \cite{bp} on 
$f_0(400-1200)$ we noted \cite{mofrascati} a relative 
suppression in comparison with $f_2(1270)$. A better study in the 1000 MeV
region is important.


\section*{Acknowledgement}
I would like to thank Peter Minkowski for the collaboration on these topics and
to Ugo Gastaldi, Stephan Narison and Hanqing Zheng for the discussion about
the scalars.

\end{document}